\documentclass{aa}

\def\amin{\ifmmode ^{\prime}\else$^{\prime}$\fi}
\newbox\grsign \setbox\grsign=\hbox{$>$}
\newdimen\grdimen \grdimen=\ht\grsign
\newbox\laxbox \newbox\gaxbox
\setbox\gaxbox=\hbox{\raise.5ex\hbox{$>$}\llap
     {\lower.5ex\hbox{$\sim$}}}\ht1=\grdimen\dp1=0pt
\setbox\laxbox=\hbox{\raise.5ex\hbox{$<$}\llap
     {\lower.5ex\hbox{$\sim$}}}\ht2=\grdimen\dp2=0pt

\begin{document}

\title{Double-periodic  blue variables in the Magellanic Clouds}

\author{R.E.\ Mennickent \inst{1}
G.\ Pietrzy{\'n}ski \inst{1}\fnmsep\inst{2}
M. Diaz\inst{3}
W.\ Gieren \inst{1}
}

\institute{Universidad de Concepci{\'o}n, Departamento de F\'{\i}sica,
 Casilla 160--C, Concepci{\'o}n, Chile\\
\email{rmennick@stars.cfm.udec.cl, pietrzyn@hubble.cfm.udec.cl, wgieren@coma.cfm.udec.cl}
\and
Warsaw University Observatory, Al. Ujazdowskie 4,00-478, Warsaw, Poland
\and
 Instituto Astron\^{o}mico e Geof\'{\i}sico, Universidade de  S\~{a}o Paulo, Brazil\\
\email{marcos@astro.iag.usp.br}
}

\date{}

\abstract{We report the discovery, based on an inspection of
the OGLE-II database, of a group of blue variables in the Magellanic
Clouds showing simultaneously
two kinds of photometric variability: a short-term
cyclic variability with typical amplitude $\Delta I \sim$ 0.05 mag and
period $P_{1}$ between 4 and 16 days and a sinusoidal, long-term cyclic
oscillation with much larger amplitude $\Delta I \sim$ 0.2 mag with
period  $P_{2}$ in the range of 150-1000 days. We find that both
periods seems to be coupled through the relationship $P_{2}$ = 35.2
$\pm$ 0.8 $P_{1}$. In general, the short term variability is
reminiscent of those shown by Algol-type binaries.
We propose that the long-term oscillation could
arise in the precession of a elliptical disc fed by a Roche-lobe filling
companion in a low mass ratio Algol system.
}

\authorrunning{Mennickent, Pietrzy\'nski, Diaz  and Gieren }
\titlerunning{Double-periodic blue variables in the Magellanic Clouds}
\maketitle

\section{Introduction}

Over the past years, the microlensing projects (OGLE, MACHO, EROS)
have monitored millions of stars in the Magellanic Clouds
and Galactic bulge for variability. The resulting huge photometric databases are very well
suited not only for microlensing studies but also for many other issues of modern
astrophysics, including the distance scale, variable stars, star clusters etc.
In particular, the OGLE-II project (Udalski, Kubiak and Szymanski, 1997),
has provided
accurate BVI measurements for about 6.5 million stars from the central
parts of the Magellanic Clouds (Udalski et al.\ 1998, 2000). Based on this same
material, a unique catalog containing about 68.000 variable stars has just been
released ({\.Z}ebru{\'n} et al.\ 2001).

One example of the scientific information that it is possible to extract
from such a catalogue is shown by the work of Mennickent et al. (2002, hereafter M02), who report the discovery of
Be star candidates in the Small Magellanic Cloud
showing unexpected photometric variations. Basically, these authors
found four types of variability in targets within the 
luminosity-colour box of 
typical Be stars: Type-1 stars showing outbursts,
Type-2 stars showing sudden magnitude jumps, Type-3 stars showing
periodic variations and Type-4 stars showing random variability.
According to these authors, possible causes for Type-1 and Type-2 stars include
blue pre-main sequence stars surrounded by thermally
unstable accretion discs and white dwarfs accreting from the 
Be star envelope in a Be+WD binary. Type-4 stars could be classical Be stars.
On the other hand, Type-3 stars could not be
linked to the Be star phenomenon at all, mainly due to their rather red colours
and strictly periodic behaviour.
The above study is complemented by the work by Keller et al. (2002) on blue
variable stars from the MACHO database in the LMC. These authors
report basically the same photometric behaviour in a sample of LMC blue
variables. These authors also find emission lines in the spectra of many
of these variables.

Further inspection of the misterious Type-3 stars found
in the SMC and the LMC (Mennickent et al. in preparation)
 revealed 30 objects showing double-periodic
photometric variations: a long-period sinusoidal variation and a short-period
modulation. 
Some of these  short-term light curves are typical of
eclipsing Algol variables.
The study of this new double-period photometric variability
is the subject of this paper.

\section{Light curve analysis and the long-term oscillation}

Table 1 lists the Type-3 stars in the SMC and LMC where two photometric periods
were found while Fig.\,1 shows some examples of light curves. MACHO light curves
were also inspected to confirm the long-term variations and search for colour variability.
MACHO identifications are also
provided along with the OGLE-II names in Table 1.
In general,
the long-term periodicity was evident from the first inspection
of the light curve. The period $P_{2}$ in these cases was calculated using
algorithms based on 
the $F$ (variance ratio) statistics and fourier transform giving
consistent results. The results shown in Table 1 reveal that
the long-term variations have periods between 140$-$960 d.
The period error was calculated as the
half width at half maximum of the periodogram peak. Trial
and error tests showed that these are upper limits for
the error and the accuracy of $P_{2}$ in Table 1 is likely
better than shown by a factor 2 at least. After finding $P_{2}$,
we applied a non-linear least squares fit to the data of the type:  \\

$I = I_{0} - A_{I} sin (2 \pi (HJD-HJD_{0})/P_{2})$ \hfill(1)  \\

where $I_{0}, A_{I}$ and $HJD_{0}$ were parameters to be determined while the period
$P_{2}$ remained fixed.  The half-amplitudes $A_{I}$ found were between 0.03 and
0.26 mag. The residuals to the best fit were examined with the same
period searching routines employed for the analysis of
the long-term oscillation. The surprising result was that in all cases listed in
Table 1 we find a second period $P_{1}$, in the range of 2$-$16 days,
which is listed along with their error
in Table 1. For non-eclipsing stars this error was found in the same way that for
$P_{2}$, but for 5 eclipsing stars the accuracy was usually
much better and the error was estimated interactively.

We also studied the MACHO data for all the variables in
our sample,
constructing the $b$, $r$ and $b-r$ light curves, and
searched for periodicities.
We confirmed the periods found on the basis of 
the OGLE-II database, and found that 
for all cases the long-term oscillation is redder at maximum (Fig.\,2). 
In general, the colour variations are of very low amplitude (Table 1).

\section{On the short-term light curves}

Once the long-period sinusoidal modulation is subtracted,
the phase curve of the residuals (Fig.\,1) 
turned out to be of three types: eclipsing, double-wave and single-wave.
The light curve of the
eclipsing stars are typical of those found in
Algol type variables. 
On the other hand, the fact that
the stars showing double-wave or single-wave light curve
share similar long-term variability
with the eclipsing stars and exhibit similar correlation
between the short-term and long-term period (see next
section) suggests that we are in presence of an homogenous group of short-period variables, 
and that these non-eclipsing stars could also be
Algol-like stars but seen under higher orbital inclinations. The
period range along with the fact that emission lines have been observed in Type-3
stars (M02 and Keller et al.\,2002) are consistent
with this interpretation. In this view,
double-wave light curves arises from the changing aspects of a
non-spherical Roche-lobe filling secondary star, as
occurs in the suspected Algol-type binary V\,1080 (\v{S}imon et al. 2000).  
It is worth to mention that  long-term  periodic
oscillations have been repported  in a few Algol systems  but with
very low amplitude (e.g. Walter 1981).
The fact that Algols
contain B-A type primaries and cooler secondaries
should explain the distribution of Type-3
stars in their colour-colour diagram (M02). 
In 5 cases we found single-wave light curves. It is possible that
the true period  is twice the reported period in these cases.
This is confirmed by the similar power shown by $P_{1}$ and 2$\times P_{1}$
in the periodogram  and also by the good fit  of $2 \times P_{1}$ by the 
$P_{1}-P_{2}$ linear correlation (see next section).

\section{The $P_{2}-P_{1}$ relationship}

In Fig.\,3 we show what we consider the most surprising result of this paper.
The $P_{2}$ and $P_{1}$ periods seems to be correlated.  All
five deviant points correspond to single-wave light curves,
so it is possible that the true orbital period has been missed by a factor 2.
The value $2 \times P_{1}$ is also indicated in the figure.
A linear least squares fit passing by zero gives:\\

$P_{2} = 35.17(75) P_{1}$\hfill(2)\\

\noindent
with a correlation coefficient 0.97 and standard deviation of 28 days.
The existence of such a  relationship suggests that the phenomenon causing the long-term
oscillation could be directly related to the binary nature  of the system.

\section{Discussion: on the nature of the long-term oscillation}

The long-term well behaved sinusoidal modulations observed in
our sample is not known to occur in other types of
blue variable stars. Here we propose as a
possible cause for the long-term oscillation the precession of an
elliptical disc around a blue star in a semi-detached binary system.
The origin of the disc ellipticity and precession
could be the tidal interaction between the disc and
the Roche-lobe filling secondary star,
in a similar way that occurs in short-orbital period dwarf novae
of the SU UMa class during superoutburst (e.g. Patterson 2001).

\begin{table*}
\caption[]{The double-periodic blue stars. OGLE name and MACHO identification are given
for every star, along with their photometric period and error.
The half amplitude of the long-term
variation is also given for the OGLE $I$ band and the MACHO $b$ band, along with the
epoch (HJD$-$2\,400\,000) for the faint-to-bright mean level crossing for the
MACHO $b$ band. The half amplitude of the MACHO $b-r$ light curve
is also given, along with their difference in phase with the $b$ curve, considering
the epoch for positive to negative mean $b-r$ level crossing as reference.
Dashes indicate non-detected or marginal colour variations. A note indicates the
appearance of the short-term variability: double-wave (dw),
single wave (sw) or eclipsing with two minima (e).}
\begin{center}
\begin{tabular}{lrrrrrrrrr} \hline  \multicolumn{1}{c}{Star}&
\multicolumn{1}{c}{MACHO ID} & \multicolumn{1}{c}{$P_{1}$ (d)} &
\multicolumn{1}{c}{$P_{2}$ (d)} &
 \multicolumn{1}{c}{$A_{I}$}   &
 \multicolumn{1}{c}{$A_{b}$}   &
  \multicolumn{1}{c}{$E_{b}$}   &
  \multicolumn{1}{c}{$A_{b-r}$}   &
   \multicolumn{1}{c}{$\Delta\Phi$}   &
\multicolumn{1}{c}{Note}\\
\hline \hline
OGLE00451755-7323436 &212.15675.158  &  5.178(5)   &  171(15)  &0.12 &0.11&50350.2(8)&0.02&-0.35& dw   \\
OGLE00474820-7319061 &212.15847.466  &  5.497(7)   &  177(10)  &0.12 &0.10&50329.3(7)&- &-& dw      \\
OGLE00553643-7313019 & 211.16304.169 &  5.092(5)   &  176(17)  &0.08  &0.04&50238(1)&0.02&0.53&   dw        \\
OGLE05025323-6909493 &  1.3686.53    &  8.025(18)  &  255(30)  &0.17  &-&-&-&-&                 dw \\
OGLE05040378-6917508 &  1.3805.130   &  6.223(7)   &  207(20)  &0.11  &-&-&-&-&                 dw  \\
OGLE05060009-6855025 &  1.4174.42    &  3.849(7)   &  230(22)  &0.19  &-&-&-&-&                 sw     \\
OGLE05101621-6854290 &  79.4900.185  &  4.301(6)   &  319(34)  &0.06  &0.04&50344(2) &0.02&0.53  & sw     \\
OGLE05115466-6846369 &  2.5144.4555  &  9.138(2)   &  361(29)  &0.26  &-&-&-&-&                 e\\
OGLE05142677-6910559 &  79.5501.400  &  6.515(2)   &  224(14)  &0.13  &0.08&50430(2)&0.04&-0.43  & e  \\
OGLE05143758-6852259 &  79.5506.139  &  5.372(6)   &  185(11)  &0.14  &0.10&50237.3(5)&0.03&0.55 &dw?   \\
OGLE05152654-6923257 &  79.5740.5092 &  6.292(8)   &  205(15)  &0.04  &0.04&50275.3(6)&-&-& dw    \\
OGLE05155332-6925581 &  79.5739.5807 &  7.2835(16) &  188(11)  &0.11  &0.08&50254(1)&0.03&0.54 &  e      \\
OGLE05171401-6936374 &  78.5979.58   &  8.309(4)   &  311(21)  &0.13  &0.08&50345(1)&0.03&0.48 &  e  \\
OGLE05194110-6931171 &  78.6343.81   &  6.9044(10) &  226(13)  &0.11  &0.06&50247(2)&0.02&0.59  & e  \\
OGLE05195898-6917013 &  80.6468.83   &  2.410(10)  &  140(5)   &0.06  &0.04&50267.6(8)&-&-  &sw       \\
OGLE05203325-6910146 &  80.6469.95   &  5.737(5)   &  182(10)  &0.04  &0.05&50357.1(8)&-&- &dw        \\
OGLE05260516-6954534 &  77.7426.140  &  3.632(3)   &  233(15)  &0.07  &0.07&50425.7(7)&-&-  &     sw       \\
OGLE05274332-6950556 &  77.7669.1013 &  7.320(11)  &  227(20)  &0.05  &0.03&50280(1)&0.02&0.54 &  dw   \\
OGLE05285370-6952194 &  77.7911.26   &  15.854(31) &  620(70)  &0.07  &0.06&50388(2)&0.01&0.67  & dw    \\
OGLE05294913-6949103 &  77.8033.140  &  7.184(8)   &  258(20)  &0.10  &0.09&50382.2(6)&0.01&-0.38& dw   \\
OGLE05295881-6934075 &  77.8036.5142 &  5.597(4)   &  179(8)   &0.12  &-&-&-&-&                 dw     \\
OGLE05313130-7012584 &  7.8269.36    &  9.231(21)  &  960(176) &0.04  &0.03&51226(4)&-&-  & sw         \\
OGLE05333926-6956229 &  81.8636.51   &  7.863(15)  &  257(18)  &0.03  &0.01&50335(2)&0.01&0.52 &  dw         \\
OGLE05371342-7010580 &  11.9237.2121 &  10.913(23) &  421(40)  &0.12  &-&-&-&-&                 dw         \\
OGLE05390681-7027487 &  11.9475.96   &  6.967(12)  &  276(15)  &0.10  &0.01&50335(2)&0.01&0.52   &dw      \\
OGLE05390992-7019262 &  11.9477.138  &  6.632(8)   &  198(15)  &0.10  &0.05&50301(1)&0.02&0.56  & dw      \\
OGLE05391746-7044019 &  11.9592.22   &  7.151(8)   &  219(20)  &0.06  &0.04&50401.0(8)&0.01&-0.49& dw        \\
OGLE05410217-7011043 &  76.9842.2444 &  7.352(13)  &  264(26)  &0.11  &0.11&50395(2)&-&-   &      dw       \\
OGLE05410942-7002215 &  76.9844.110  &  6.586(9)   &  245(23)  &0.12  &0.09&50510(2)&-&-     &    dw         \\
OGLE05435003-7057431 &  15.10314.144 &  5.012(5)   &  173(12)  &0.12  &0.09&50250.7(7)&-&-    &   dw         \\
\hline
\end{tabular}
\end{center}
\end{table*}

The 3:1 resonance between a disc particle orbiting the primary and
the binary system occurs at $R_{disc} \approx 0.46a$, where
$a$ is the binary separation. Elliptical orbits at this radius will experience a
dynamical apsidal advance with a period given by:\\

$P_{p} = \frac{\sqrt{1+q}}{0.37q}(R_{disc}/0.46a)^{-2.3}P_{o}$ \hfill(3)\\

(Murray 2000) where $q$ is the ratio between the mass of the
secondary star and the primary star and $P_{o}$ the orbital period.
If we interpret $P_{1}$ as the orbital period  and $P_{2}$ as the precession period, then the above equation
imposes a strong correlation between both periodicities,
as effectively seen in Fig.\,3.
The observed slope of $P_{2}/P_{1}$ $\approx$ 35 should imply a mass ratio
$q \approx$ 0.08 for $R_{disc} = 0.46 a$. There are few Algols
with $q <$ 0.1 (e.g. Richards \& Albright 1999), and they do not show
the long-term oscillations described in this paper. However it is
exciting that only in these low $q$ systems two important conditions
are fulfilled:
the primary radius is smaller than the stream circularization radius,
being possible the formation of
an accretion disc (e.g. Richards \& Albright 1999) {\it and} the 3:1 resonance radius
is below the tidal truncation radius, so the disc could in principle
grown beyond the 3:1 resonance and may eventually start to precess (Fig.\,4).
Long-term spectroscopic observations and detailed modeling are required to
confirm this scenario.

\begin{acknowledgements}

REM acknowledges support by Grant Fondecyt 1000324 and
DI UdeC 202.011.030-1.0. MPD thanks CNPq support under grant
\# 301029. We
thank the OGLE collaboration team for making their data
public domain.
This paper also utilizes public domain data originally obtained by the
MACHO Project.

\end{acknowledgements}

{\bf Figure Captions}

Fig. 1: Eclipsing (upper graph) and double-wave (below graph)
light curves. The three pannels in every graph show, from bottom to top,
the long-term light variability, the residuals from a sinusoidal fitting
and the residuals folded using the short-term period. 

Fig. 2: MACHO $b$ and $b-r$  light curves folded with the long-term period.

Fig. 3: The long-term period $P_{2}$ versus the short-term period $P_{1}$. 
The errors in  $P_{1}$ are smaller than the used symbols.
Dashed horizontal lines connect two possible solutions  for
stars with single-wave short-term light curve. The linear fit given by Eq.(2)
is also shown.

Fig. 4: The primary radius in units of the the binary separation for the Algols
shown by Richards \& Albright 1999 (filled circles). The circularization radius,
the tidal truncation radius and the 3:1 resonance radius, all of them in units of the
binary separation, are shown as a function
of the mass ratio. Theoretical expressions for these quantities 
have been taken from Warner (1995).
It is possible that only Algols with very low mass ratios could
maintain (precessing) accretion discs beyond the 3:1 resonance.

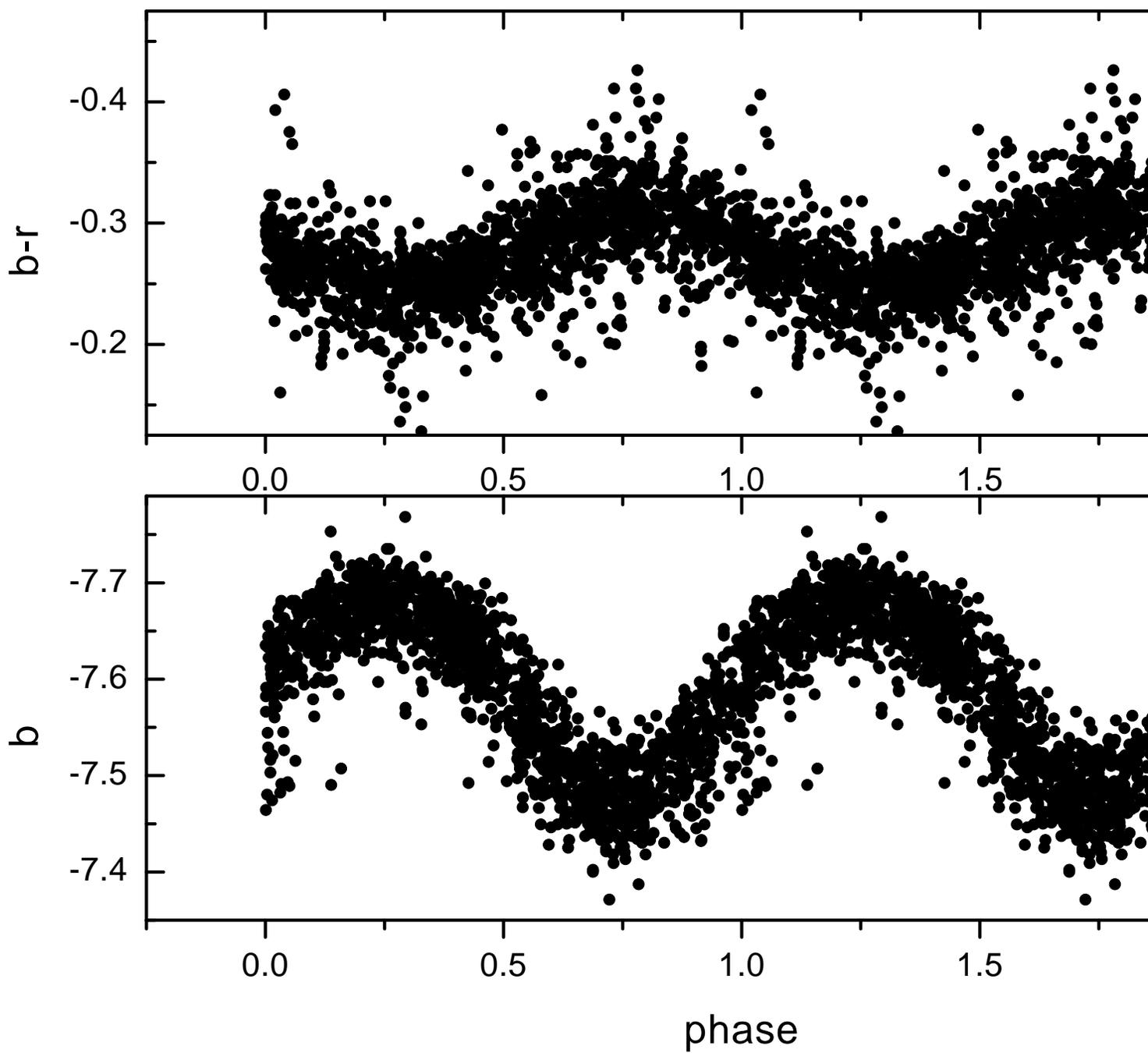
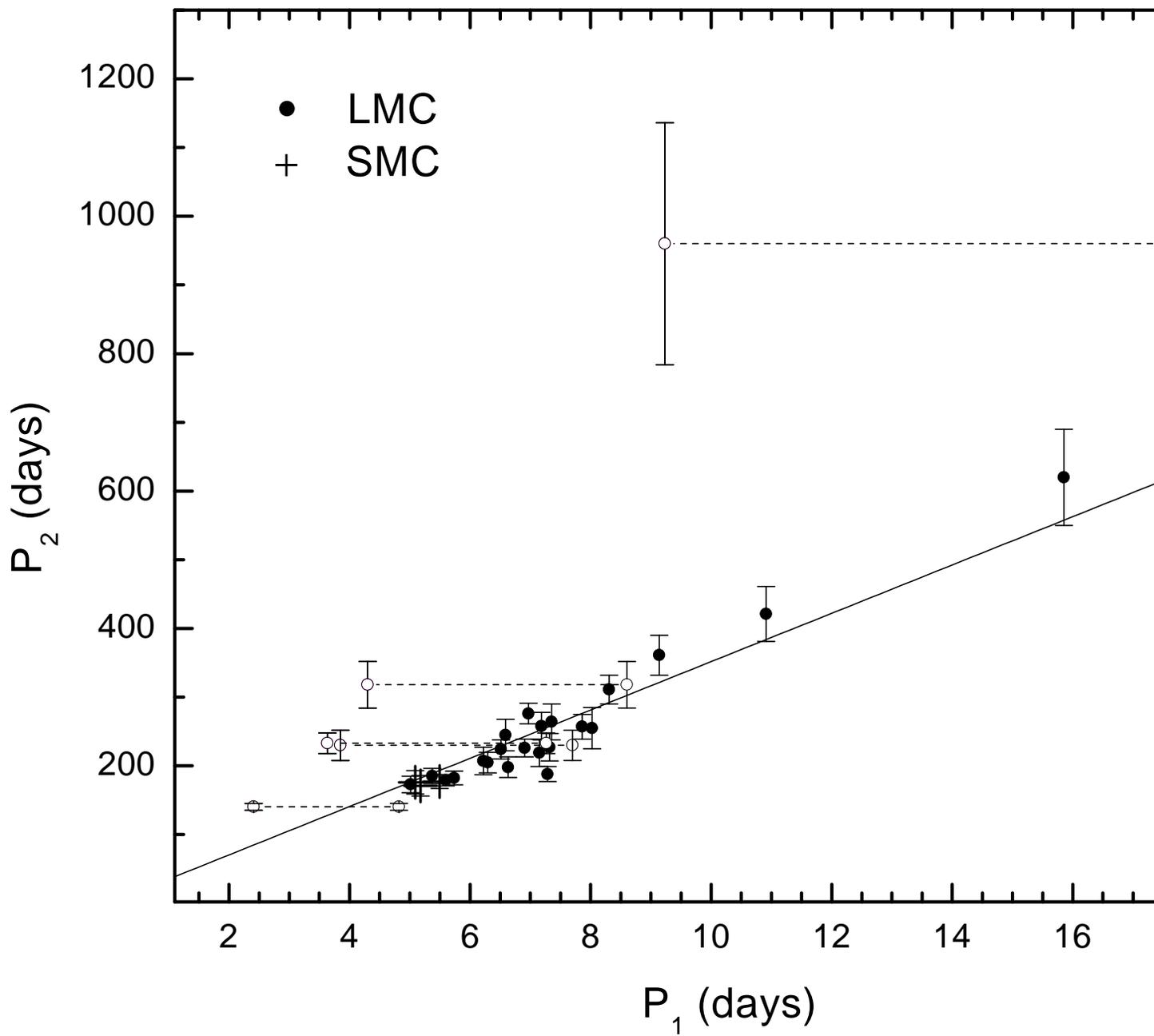
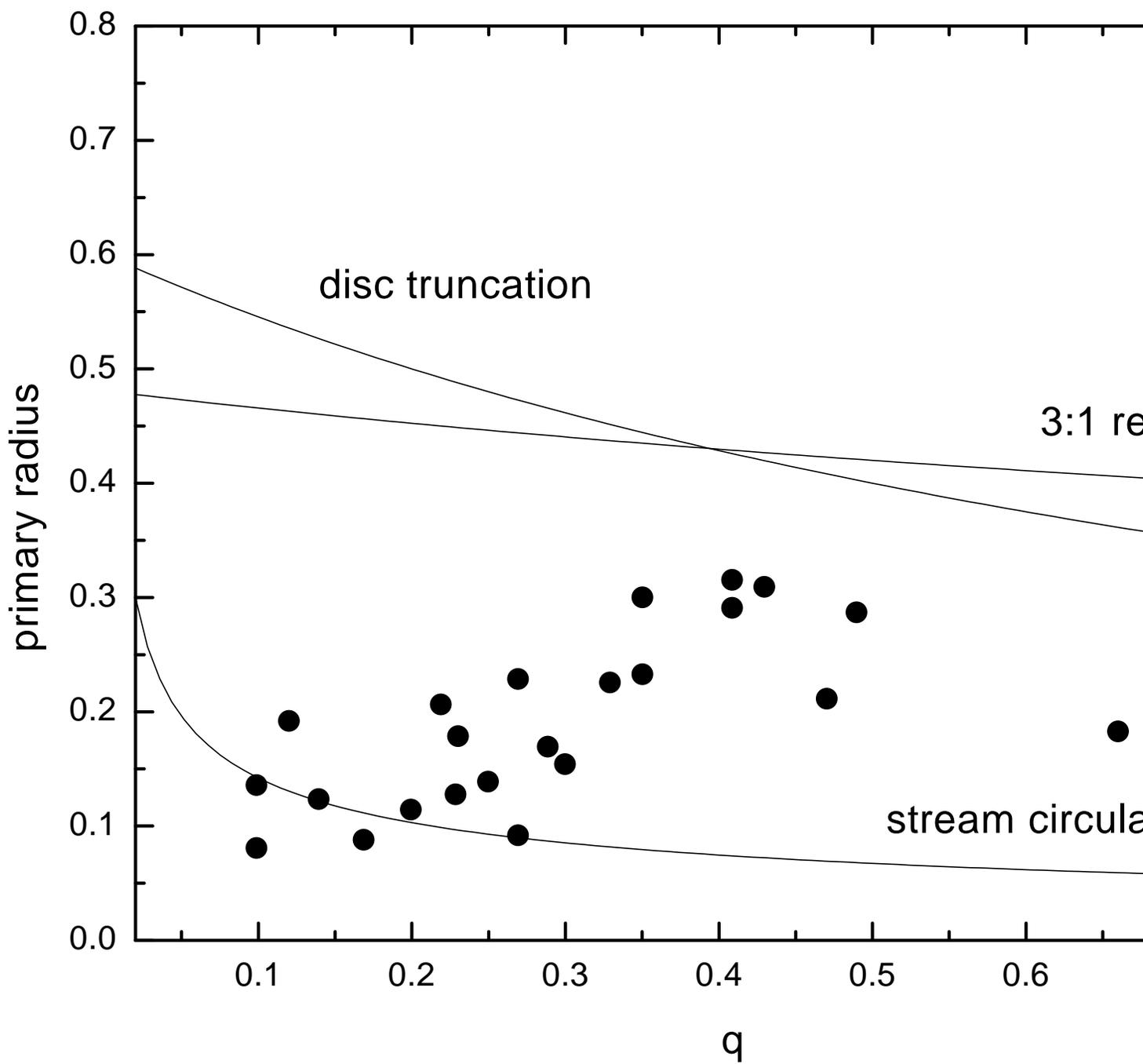
\end{document}